\documentclass[12pt,preprint]{aastex}

\newcommand{\Msun}{\ensuremath{\mathrm{M}_\sun}}

\newcommand{\nuc}[2]{\ensuremath{\mathrm{^{#1}#2}}}
\newcommand{\FLASH}{{\sc flash}}
\begin{document} 

\title{Initiation of the detonation in the gravitationally confined detonation model of Type Ia supernovae}

\author{
Ivo R. Seitenzahl\altaffilmark{1,2,3}, Casey A. Meakin\altaffilmark{2,4,5,6}, Don Q. Lamb\altaffilmark{4,5}, James W. Truran\altaffilmark{2,4,5,7,8}
}
\altaffiltext{1}{Department of Physics,
                 The University of Chicago,
                 Chicago, IL  60637}
\altaffiltext{2}{Joint Institute for Nuclear Astrophysics,
                 The University of Chicago,
                 Chicago, IL  60637}
\altaffiltext{3}{Max-Planck-Institute for Astrophysics,
                 85741 Garching, Germany}
\altaffiltext{4}{The Center for Astrophysical Thermonuclear Flashes,
                 The University of Chicago,
                 Chicago, IL  60637} 
\altaffiltext{5}{Department of Astronomy and Astrophysics,
                 The University of Chicago,
                 Chicago, IL  60637}	
\altaffiltext{6}{Steward Observatory,
                 The University of Arizona,
                 Tucson, AZ  85719}	
\altaffiltext{7}{Enrico Fermi Institute,
                 The University of Chicago,
                 Chicago, IL  60637}
\altaffiltext{8}{Argonne National Laboratory,
                 Argonne, IL 60439}

\begin{abstract}

We study the initiation of the detonation in the gravitationally
confined detonation (GCD) model of Type Ia supernovae (SNe Ia).  In
this model, ignition occurs at one or several off-center points,
resulting in a burning bubble of hot ash that rises rapidly, breaks
through the surface of the star, and collides at a point on the
stellar surface opposite the breakout, producing a high-velocity
inwardly directed flow. Initiation of the detonation occurs spontaneously in a region where
the length scale of the temperature gradient extending from the flow
(in which carbon burning is already occurring) into unburned fuel is
commensurate to the range of critical length scales which have been
derived from 1D simulations that resolve the initiation of a
detonation. By increasing the maximum resolution in a truncated cone that encompasses this region, beginning somewhat before initiation of the detonation occurs, we successfully simulate in situ the first gradient-initiated detonation in a whole-star simulation.  The detonation emerges when a compression wave overruns a pocket of fuel situated in a Kelvin-Helmholtz cusp at the leading edge of the inwardly directed jet of burning carbon.  The compression wave pre-conditions the temperature in the fuel in such a way that the Zel'dovich gradient mechanism can operate and a detonation ensues.  We explore the dependence of the length scale of the temperature gradient on spatial resolution and discuss the implications for the robustness of this detonation mechanism.  We find that the time and the location at which initiation of the detonation occurs varies with resolution.  In particular, initiation of a detonation had not yet occurred in our highest resolution simulation by the time we ended the simulation because of the computational demand it required.  However, it may detonate later.  We suggest that the turbulent shear layer surrounding the inwardly directed jet provides the most favorable physical conditions, and therefore the most likely location, for initiation of a detonation in the GCD model. 

\end{abstract}

\section{Introduction}
\label{sec:intro}

Type Ia supernovae (SNe Ia) are among the most luminous optical transients.
They contribute significantly to the non-primordial elements in the universe, especially iron \citep{timmes95}.
A large fraction of SNe Ia exhibit a correlation between peak luminosity and rate of decline that can be used to make them standard candles, and therefore excellent distance indicators and probes of cosmology [see, e.g. the reviews by \citet{leibundgut01} and \citet{frieman08}]. 
While SNe Ia are thought to be the result of the thermonuclear incineration of white dwarfs (WDs) \citep{hoyle60}, the explosion mechanism is not fully understood [see e.g., the reviews by \citet{hillebrandt00} and \citet{podsiadlowski08}].

Most astronomers favor a scenario in which the exploding star is a single, near Chandrasekhar-mass WD.
Consequently, multi-dimensional full-star simulations of the explosion of near Chandrasekhar-mass CO white dwarfs have been performed by several research groups. 
These models posit the ignition of a turbulent deflagration flame in the core of the WD \citep[e.g.][]{garcia-senz95,hoeflich02,kuhlen06},
which eventually gives rise to a supersonic detonation wave. 

In the deflagration to detonation transition (DDT) model, the
detonation is postulated to occur when the deflagration flame reaches
densities low enough that the Gibson length $l_G$, defined as the
scale where the turbulent velocities equal the laminar flame speed,
becomes smaller than the flame width $\delta$ \citep{khokhlov97,niemeyer97}.  When this criterion is met, turbulence tears the flame apart, producing distributed burning.
Simulations of SNe Ia that explode via DDT have been extensively studied in multiple dimensions \citep[e.g.][]{arnett94a,arnett94b,livne99,gamezo04,gamezo05,golombek05,roepke07c,bravo08}. 

The pulsational delayed detonation model (PDD) posits a centrally ignited deflagration, which undergoes a deflagration to detonation transition after one or several pulsations \citep{ivanova74,khokhlov91c}. 
A variant of this model is the pulsational reverse detonation (PRD) model, in which the initial deflagration releases energy and expands the star but quenches before it becomes gravitationally unbound.  The detonation in these models is postulated to form from an accretion shock during the re-collapse of the star after the deflagration ash has risen and fresh fuel occupies again the center of the WD \citep{dunina-barkovskaya01,bravo05,bravo06}.  

In the gravitationally confined detonation model (GCD), the
deflagration flame ignites at one or more points slightly offset from
the center of the WD
\citep{plewa04,plewa07,roepke07,townsley07,jordan08,meakin09}.  As the
bubble grows and rises, the Rayleigh-Taylor instability causes rapid
growth of the flame surface resulting in a rate of nuclear burning far exceeding the rate that would occur due to laminar burning alone  \citep[e.g.][]{khokhlov95,zhang07}. 
The bubble quickly accelerates, burning on the order of a few percent of the mass of the star by the time it breaks through the surface of the WD \citep[e.g.][]{livne05}. 
Owing to the large surface gravity of the WD, the bulk of the ash in the bubble remains gravitationally bound.  The cold ash flows over the surface of the star and converge at the opposite point on the stellar surface from where breakout occurred.  The converging flow of cold ash pushes unburned fuel in the surface layers of the WD ahead of it, compressing it, and increasing its temperature and pressure enough that the carbon in it begins to burn.  Inwardly and outwardly directed jets form at the stagnation point, driving the inwardly directed jet to high densities [ see the discussion in \citet{meakin09}].
The inwardly directed jet is unstable to shear instabilities and Kelvin-Helmholtz rolls are observed to form. 
Additionally, strong sound waves originating from the surface flow  traverse the jet.

It is in this environment that the detonation in the GCD model is thought to occur.  To determine whether initiation of a detonation occurs under these conditions is, however not straightforward, primarily because the length scales pertinent to the detonation initiation are typically much smaller than the highest resolution possible in 2D, and even more so in 3D, whole-star simulations. 
In this context, past simulations of the GCD scenario can be grouped into two categories:
\begin{enumerate}
\item Those that limit nuclear burning to the flame model and don't simulate the detonation phase, including initiation of the detonation  \citep{roepke07,townsley07,jordan08}.
\item Those in which the detonation emerges in situ through nuclear burning outside the flame model \citep{plewa04,plewa07,meakin09}.
\end{enumerate}

Simulations in the first category only infer the successful initiation (or failure) of the detonation based on a comparison of the temperatures and densities reached in the full-star simulation with those for the ``critical detonation conditions'' derived from 1D reactive hydrodynamics simulations \citep[e.g.][]{niemeyer97} at higher resolution.  
This procedure is inherently uncertain, due to the sensitivity of the detonation conditions to the functional form of the unresolved and therefore unknown temperature profile \citep{seitenzahl09b}.
The practice of using the temperatures and corresponding densities in a simulation without distributed burning (i.e. in which nuclear burning is not allowed to occur outside the flame model) suffers from additional limitations. 
First, it ignores the feedback of nuclear burning on the hydrodynamics. 
The energy liberated during carbon burning in the jet alters the flow and may produce a temperature structure more conducive for initiation of a detonation. 
Second, not keeping track of nuclear reactions takes away the ability to decide whether the temperatures and densities obtained in the purely hydrodynamical flow in fact occur in detonatable fuel, or the fuel has already been depleted due to the nuclear burning that occurs during the process of compressional heating. 

It is our belief that direct initiation of a detonation is hard to achieve in a CO WD environment \citep{nomoto76,mazurek77} and that spontaneous initiation via the gradient (SWACER) mechanism \citep{zeldovich70,lee78} is more likely \citep{blinnikov86,blinnikov87}.  
GCD simulations in the second category allow for the detonation to form explicitly in situ, and have paid some attention to the details of the initiation mechanism. 
However, no GCD simulations to date have spatially resolved the temperature gradients that are key to initiation of the detonation. 

In section~\ref{sec:sim} we describe the setup for the simulations reported in this paper, and the method we use to resolve the temperature gradients in the region where the detonation is expected to occur and so shed more light onto the details of the detonation mechanism in GCD.  
In section~\ref{sec:detinit} we present our results for a set of four simulations with different maximum resolutions.   
Section~\ref{sec:dis} compares our results with earlier work and discusses the implications of our results.  Section~\ref{sec:con} summarizes our conclusions.

\section{Simulations}
\label{sec:sim}
\begin{figure}
\plotone{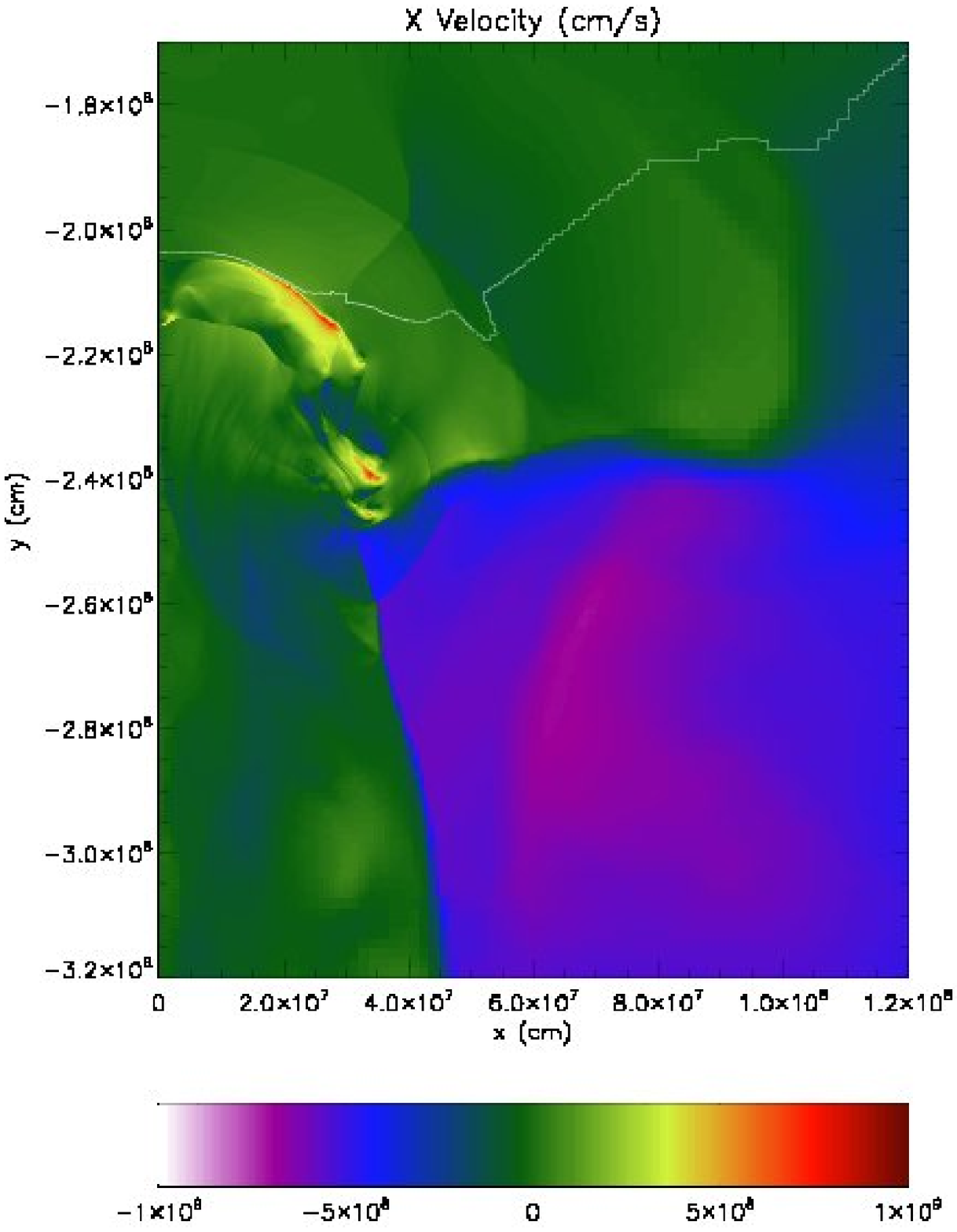}
\caption{500 m resolution. White line is $\rho = 7\times 10^6$ g cm$^{-3}$ iso-density contour. 
X-velocity of material showing matter converging at the South Pole at time $t=2.2124$ s, 54 ms before detonation. 
The accretion shock is clearly visible near $x = 400$ km off axis in this plot.}
\label{fig:500m_velx}
\end{figure}

\begin{figure}
\plotone{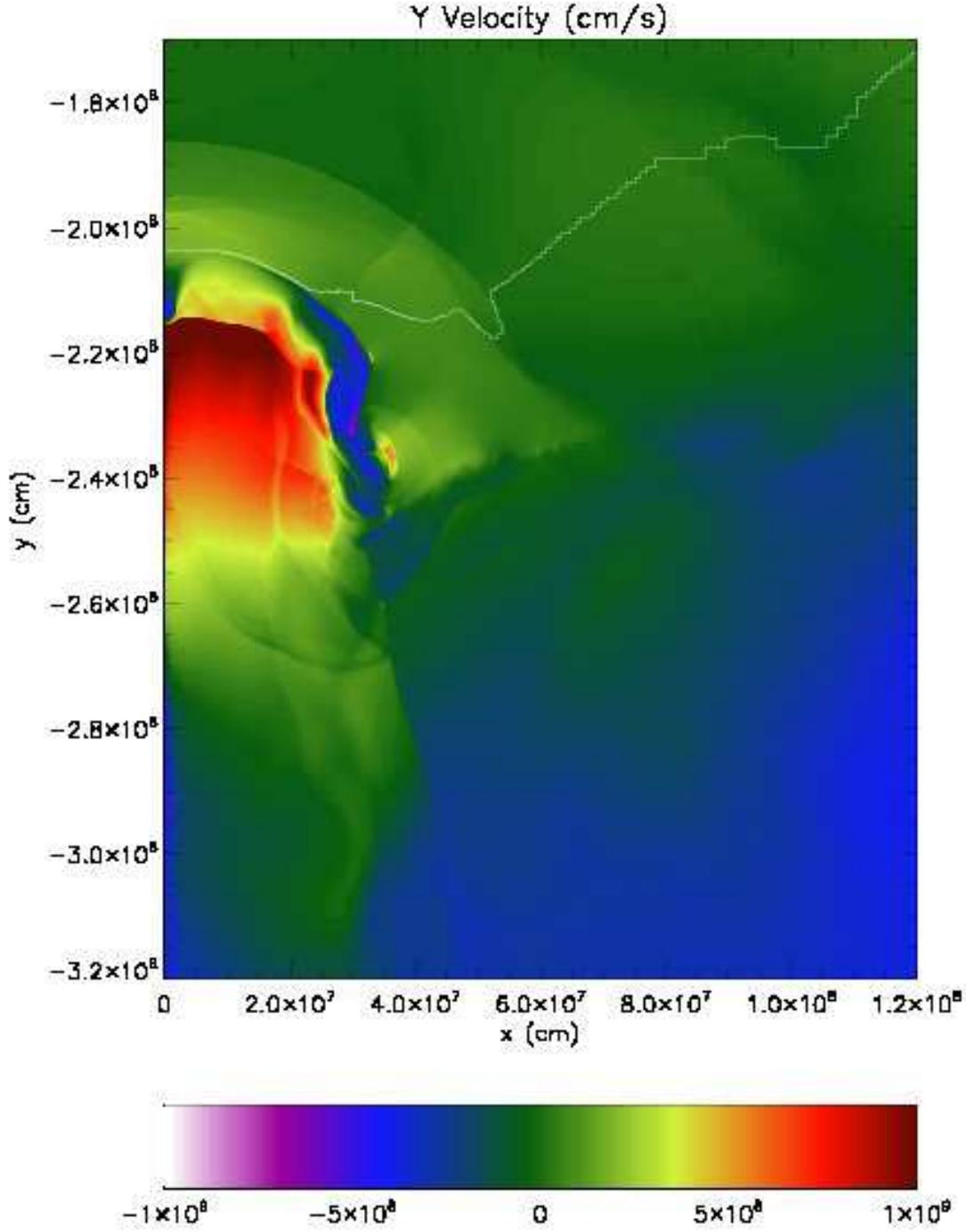}
\caption{500 m resolution. White line is $\rho = 7\times 10^6$ g cm$^{-3}$ iso-density contour.  Y-velocity of material showing jet penetrating inward 54 ms before detonation. The internal shock see near $y = -2200$ km is not the location of the initiation.}
\label{fig:500m_vely}
\end{figure}

In this paper we present 2D axisymmetric GCD simulations performed with the \FLASH\ Code.  
The way in which we carry out the simulations is identical to \citet{meakin09}, who extend the \citet{townsley07} deflagration models to the detonation phase and use the energetics of \citet{calder07} and \citet{seitenzahl09a}.  
The initial model is a cold ($T=4\times10^7$ K), isothermal WD with a mass of $1.365\; \Msun$, which has a central density of $\rho_c = 2.2\times10^9$ g cm$^{-3}$.  
The composition throughout the WD is equal parts \nuc{12}{C} and \nuc{16}{O} by mass.  
The simulations start with a single ignition bubble with radius 16 km offset from the center of the star by 40 km offset. 
In this single ignition point scenario, the center of the bubble and the center of the star define a symmetry axis.  
The bubble rises as it grows and breaks through the surface of the star at what we will refer to as the "North Pole."  
Most of the ash in the bubble remains gravitationally bound to the white dwarf.  Now rarefied and cold, the ash spreads out from the North Pole and flows over the stellar surface, compressing unburned fuel near the the opposite point on the stellar surface (which we will refer to as the "South Pole").  A conical accretion shock forms 300-400 km away from the symmetry axis (see Fig. \ref{fig:500m_velx}). 
The pressure in the compressed region exceeds the hydrostatic pressure in both radial directions, producing outwardly and inwardly directed jets (see Fig. \ref{fig:500m_vely}).

\subsection{Jet morphology}

\begin{figure}
\plotone{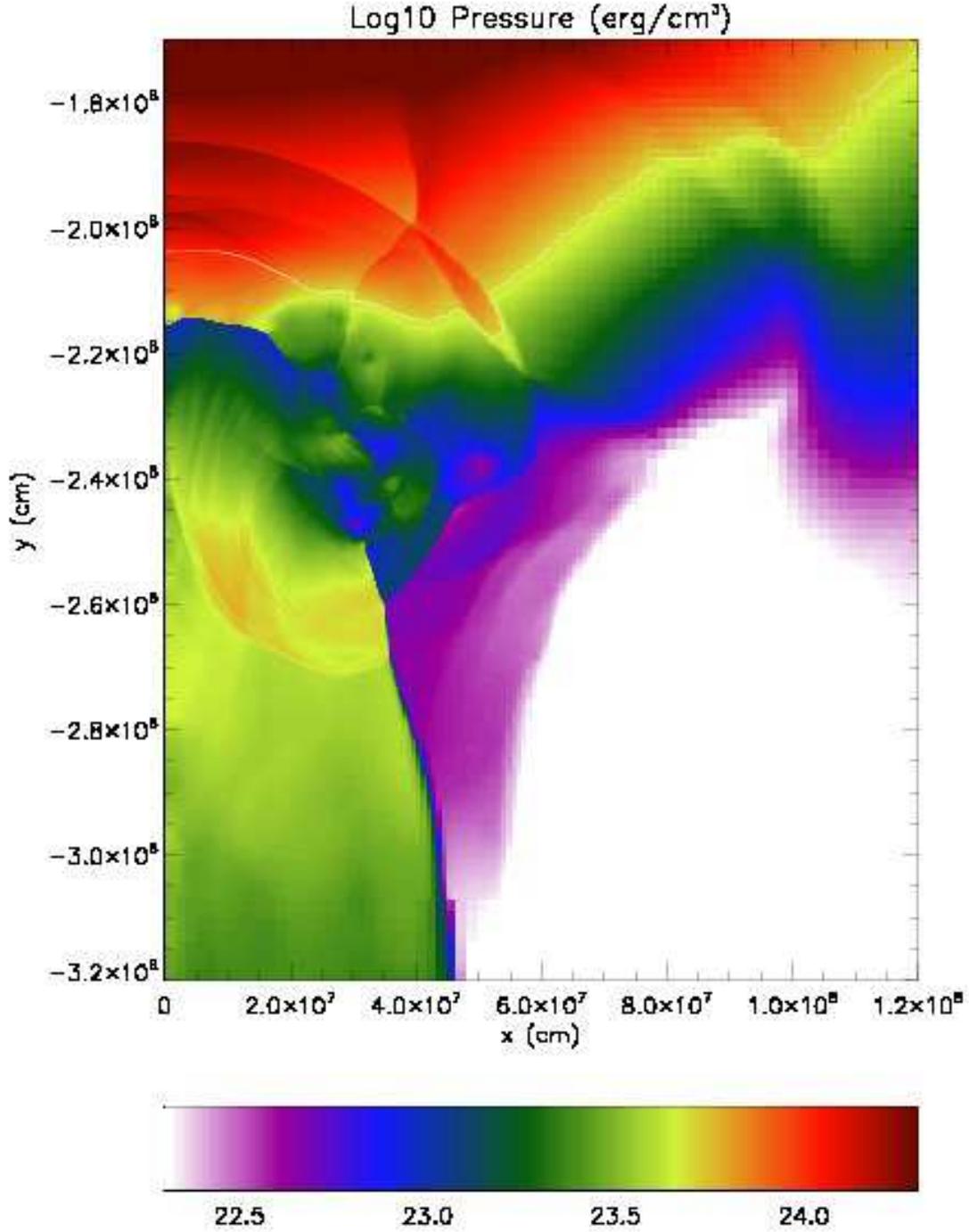} 
\caption{500 m resolution. White line is $\rho = 7\times 10^6$ g cm$^{-3}$ iso-density contour. In this pressure plot of the collision region, 54 ms before detonation, the anatomy of the jet, including the various shocks and hydrodynamic waves, is clearly visible.}
\label{fig:500m_pres}
\end{figure}

The hydrodynamical flow in the collision region is complex:
Compression waves emanate from the head of the inwardly moving jet into the star; the (locally) supersonic flow in the jet gives rise to internal shocks; strong sound waves from the interaction of the surface flow with the underlying star traverse the South Pole region (see \citet{plewa07} and Fig. \ref{fig:500m_pres} ).  
Furthermore, the boundary between the jet and the surrounding material is unstable to the generation of turbulence via the Kelvin-Helmholtz (K-H) instability and ``rolls'' characteristic of the K-H instability form and broaden the interface. 
It is in this dynamic, hot, and turbulent environment that conditions may occur that lead to the initiation of a detonation.  

\begin{figure}
\plotone{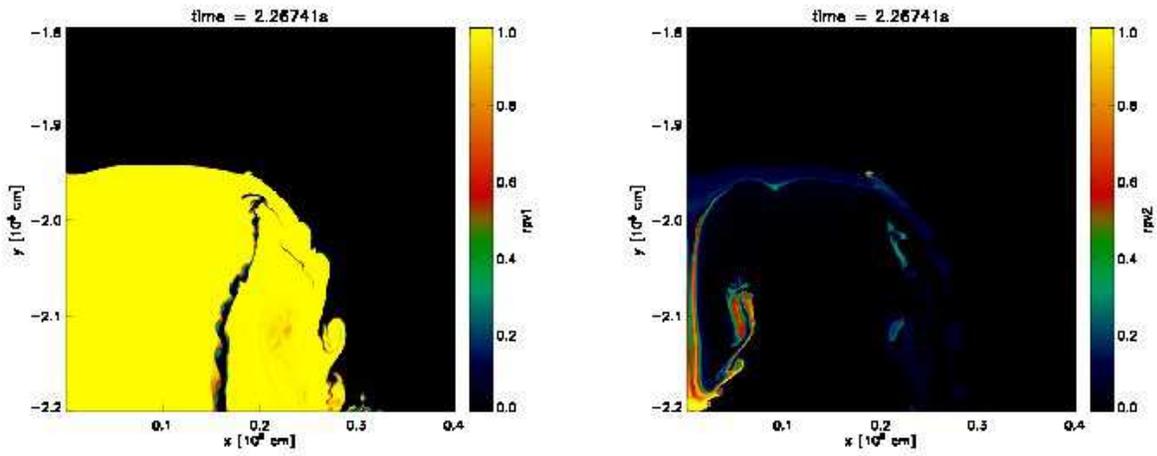}
\caption{500 m resolution. Left: Reaction progress variable $\phi_1$. Right: Reaction progress variable $\phi_2$,a tracer for the progress of the burning to Si-group elements \citep{townsley07}. Barely visible (near the center of the plots) is the detonation emerging at the boundary separating the carbon depleted jet region and the unprocessed fuel.\label{fig:500m_large}} 
\end{figure}

At the accretion shock, the pressure of the shocked material is balanced by the ram-pressure of the inflowing material \citep[c.f.][]{meakin09}. 
Compression of the fuel heats it to temperatures high enough for carbon burning to occur.  This raises the pressure even further above the hydrostatic value.  
Save for a narrow channel of cold material, carbon, as measured by the progress variable $\phi_1$ \citep[c.f.][]{townsley07}, has been depleted in the inwardly moving jet (see fig. \ref{fig:500m_large} left).  
On the other hand, oxygen  (as measured by the progress variable $\phi_2$ \citep[c.f.][]{townsley07}), is only locally depleted (mostly near the symmetry axis) and remains largely unburned in the jet (see Fig. \ref{fig:500m_large}, right panel). 
The hot material in the jet is driven inward up the density gradient while continuing to consuming carbon.

\subsection{Higher resolution truncated cone}
A goal of this paper is to determine whether the conditions near the head of the inwardly directed jet in the GCD model are conducive to the initiation of a detonation via the Zel'dovich gradient mechanism or another mechanism (e.g., SDT; see the discussion in section \ref{sec:dis}).  One-dimensional simulations show that, while the critical length scale for spontaneous initiation of a detonation depends on the temperature profile, at a density of $1.0\times10^7$ g cm$^{-3}$ it is on the order of a few km \citep{seitenzahl09b}.  Therefore, simulations with a resolution at least this fine are necessary in order to determine whether initiation of a detonation occurs via the gradient mechanism

In spite of performing our calculations only in 2D and making use of the adaptive mesh refinement (AMR) capabilities of the \FLASH\ code, full-star simulations at sub-km resolution were prohibitively expensive given the computational resources available to us.  We therefore decided on a compromise, and introduced a higher resolution truncated cone covering the region of the jet where the detonation is expected to occur.  By spatially constraining the highest resolution volume in this way, we limited the total number of computational cells while retaining the capability to resolve the length scales on which critical temperature gradients were expected to occur.

\begin{figure}
\plotone{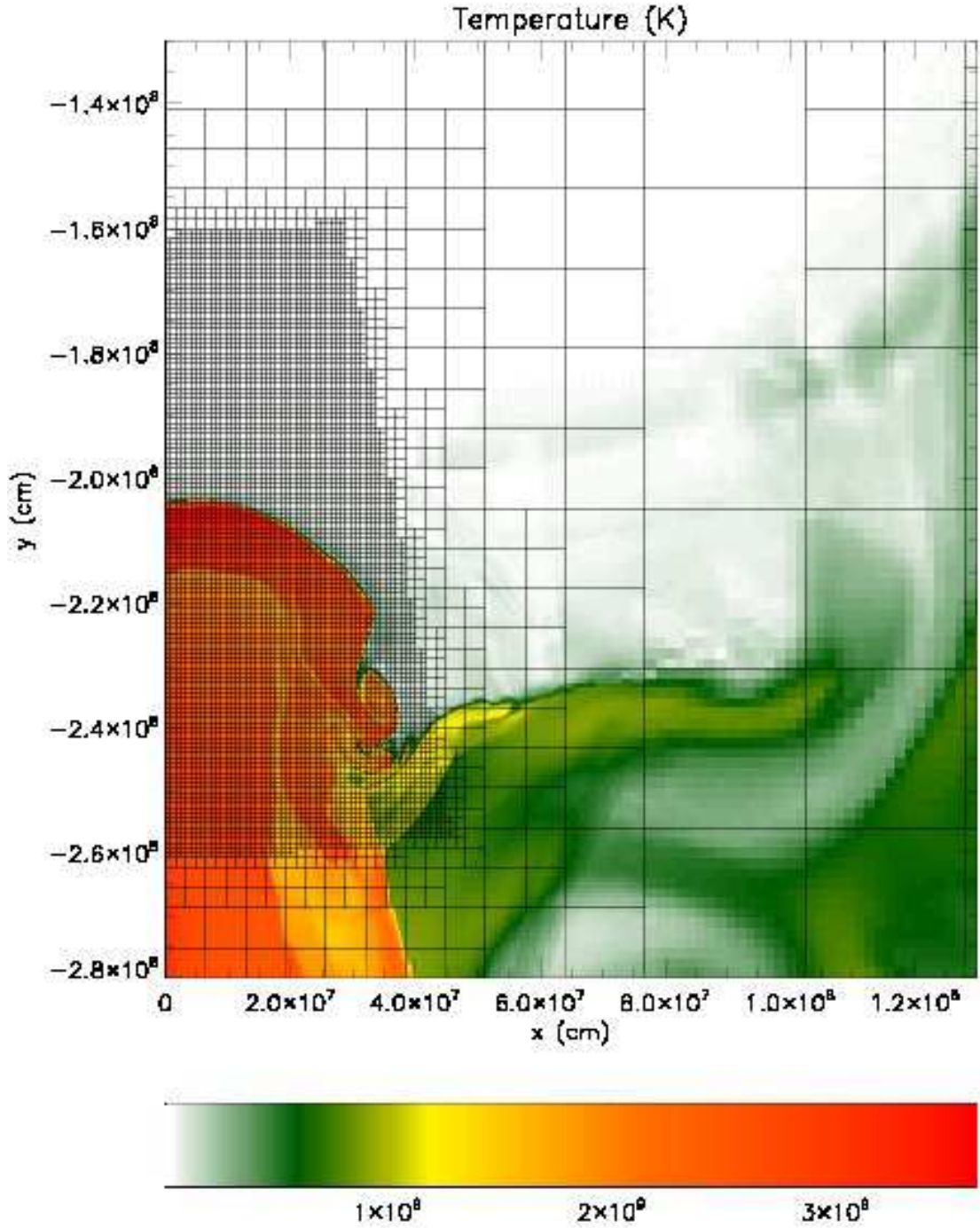} 
\caption{500 m resolution. Blocks of fine resolution ($16 \times 16$ zones per block, 500 m per zone) are used to cover a conical region near the tip of the inward moving jet, 
here visible as the region of high temperature resulting from the shock heating and subsequent carbon burning, 54 ms before detonation.}
\label{fig:500m_temp}
\end{figure}

The deflagration phase of all of the simulations presented here was identical and performed at 4 km resolution up to $t = 2.2124$ s.  At this time a higher resolution truncated cone with (half) opening angle $\alpha \approx 10^{\circ}$ was imposed on a sub-domain extending in radius from 1600 km to 2600 km (see fig. \ref{fig:500m_temp}). We chose the parameters of the highly resolved region so that it covered the compression region ahead of the inwardly moving jet at the South Pole where we expected initiation of the detonation to occur, assuring that this region was maximally refined.  We carried out four simulations: one in which the resolution in the collision region ahead of the inwardly moving jet remained 4 km, and three others in which the maximum resolution in this region was 500 m, 250 m and 125 m.




\section{Results}
\label{sec:detinit}
All but the highest resolution case (125 m) had detonated by the physical time at which we stopped the simulations.  Table~\ref{tab:conditions} gives the physical conditions (i.e., the time, location, peak and background temperatures, and fuel density) in the environment where initiation of the detonation wave occurs in the different cases.  In the 500 m and 250 m cases, initiation of the detonation is triggered by the interaction of a hydrodynamical compression wave with the leading edge of the jet.  This makes the definition of fuel density somewhat ambiguous, as one could refer to the state on either side of the compression front.  Table \ref{tab:conditions} therefore lists both the uncompressed fuel density (left) and the compressed fuel density (right) in these two cases. 

All of the detonations that occur in this study are due to the gradient mechanism; i.e., they emerge from a gradient in induction times.  We calculate the isochoric induction time, $\tau_{i,cv}(\rho_8,T_9,X_0^c)$, defined here as the amount of time that passes until $90\%
$ of the initial mass fraction of \nuc{12}{C}, $X_0^c$, is consumed in a constant volume self heating network calculation, using the formula,
\begin{equation}
\label{eq:tauicv}
\tau_{i,cv} = 1.18\times10^{-5} \; \mathrm{s} \; (X_0^C \rho_8)^{-1.85} (T_9-0.206)^{-7.7}[1+1178(T_9-0.206)^{-7.7}].
\end{equation}
This expression is a fit made by \citet{dursi06} to a large number of 13-nuclide $\alpha$-chain nuclear reaction network calculations, which the authors found was accurate to better than a factor of 5 for the densities and temperatures of interest here.  For an in depth discussion of the spontaneous initiation of detonations via the gradient mechanism and the critical length scale for detonation, see \citet{seitenzahl09b}.

Since the mechanism(s) leading to initiation of a detonation vary in the four cases we carried out, we discuss them individually.

\subsection{4.0 km resolution case}
\begin{figure}
\plotone{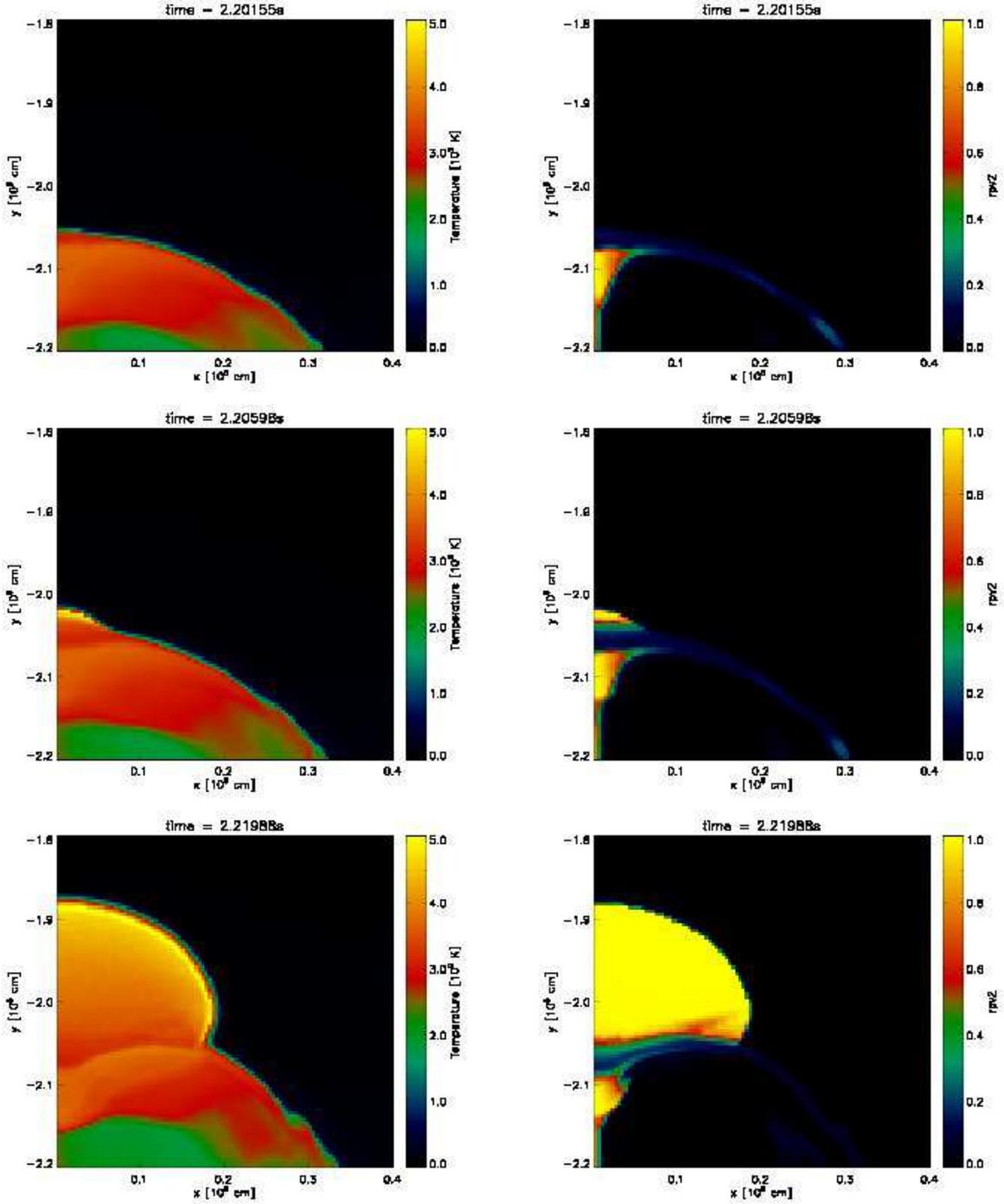}
\caption{4.0 km resolution. Left column shows a time sequence of the temperature. Right column shows reaction progress variable $\phi_2$, a tracer for the progress of the burning to Si-group elements \citep{townsley07}. \label{fig:4km_large}} 
\end{figure}

In this case (which is the lowest resolution simulation we carried out), the detonation wave occurs on the symmetry axis (see Fig. \ref{fig:4km_large}) when the head of the jet reaches a fuel density of $\rho \approx 1.1\times10^7$ g cm$^{-3}$.  It occurs $\approx 0.06$ s earlier than in the 500 m and 250 m cases discussed below (see Table \ref{tab:conditions}).  The detonation wave spreads radially outward from the initiation site.  Ignition of oxygen and the products of carbon burning occurs concurrently with initiation of the the detonation wave (see fig. \ref{fig:4km_large} right).  

Initiation of the detonation occurs via the gradient mechanism as a result of the temperature gradient that extends into unburned fuel ahead of the jet, which consists of carbon burning ash and is hot.  The length scale of the temperature gradient is $\sim 3-4$ computational zones, the number of zones over which \FLASH\ spreads the temperature jump at the head of the jet, and thus remains constant.  Since the critical length scale for initiation of a detonation decreases as the density increases, the detonation occurs when the density becomes high enough that the length scale of the induction time gradient (which for the approximately constant density conditions that exist locally in the unburned fuel is in one-to-one correspondence with the temperature gradient), becomes smaller than the critical length scale for initiation of a detonation.  We consider the initiation of the detonation in this case to be an artifact of the low resolution and numerical diffusion, which lead to a much shallower temperature gradient, and hence larger pre-heated region, than in the higher resolution cases.

\subsection{500 m resolution case}
\begin{figure}
\plotone{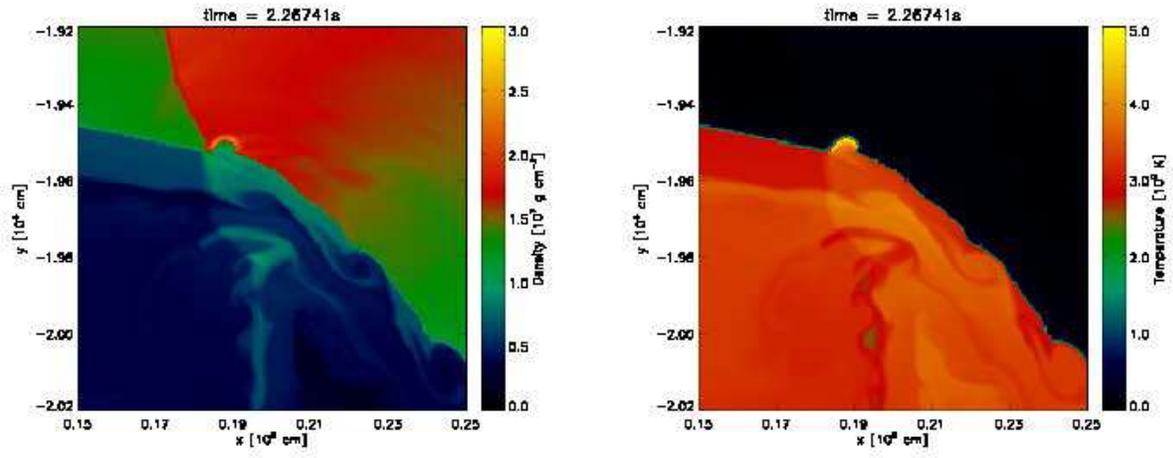}
\caption{500 m resolution. Left: Detonation forms in compressed material. Right: Temperature in bulk of compressed region is not high enough for direct initiation of detonation. \label{fig:500m_medium}} 
\end{figure}

In this case, the detonation occurs off axis (see Table~\ref{tab:conditions} and Figs.~\ref{fig:500m_large}~and~\ref{fig:500m_medium}).  It occurs $\approx 0.06$ s later than in the 4 km case and at a density of $\sim\rho=1.2\times10^7$ g cm$^{-3}$, which is higher than in the 4 km case.  The reason is that the length scale of the temperature gradients at the leading edge of the jet are still $\sim 3-4$ computational zones; i.e., the are an eighth of what they were in the 4 km case.  Consequently, the temperature gradient at the time and location where the detonation was launched in the 4 km case are now too steep.  However, the critical temperature gradient is steeper at high densities (see Table 4 in \citet{seitenzahl09b}).  Detonation at the higher resolution is merely delayed until the jet has reached densities high enough that
the critical length scale is satisfied.

The detonation emerges behind a compression front that moves more or less transverse to the tip of the jet (see Fig. \ref{fig:500m_medium} left).  The compression wave is not strong enough, i.e., the temperature behind the wave is not high enough (see Fig. \ref{fig:500m_medium} right), for direct initiation of a detonation.  The wave, however, plays a crucial role in the initiation.  It compresses and heats the carbon in the transition layer at the head of the jet, leaving behind an induction time gradient which eventually becomes shallow enough to lead to a 
detonation as the jet moves to higher density (see Fig. \ref{fig:tiny}).

\subsection{250 m resolution case}
\begin{figure}
\plotone{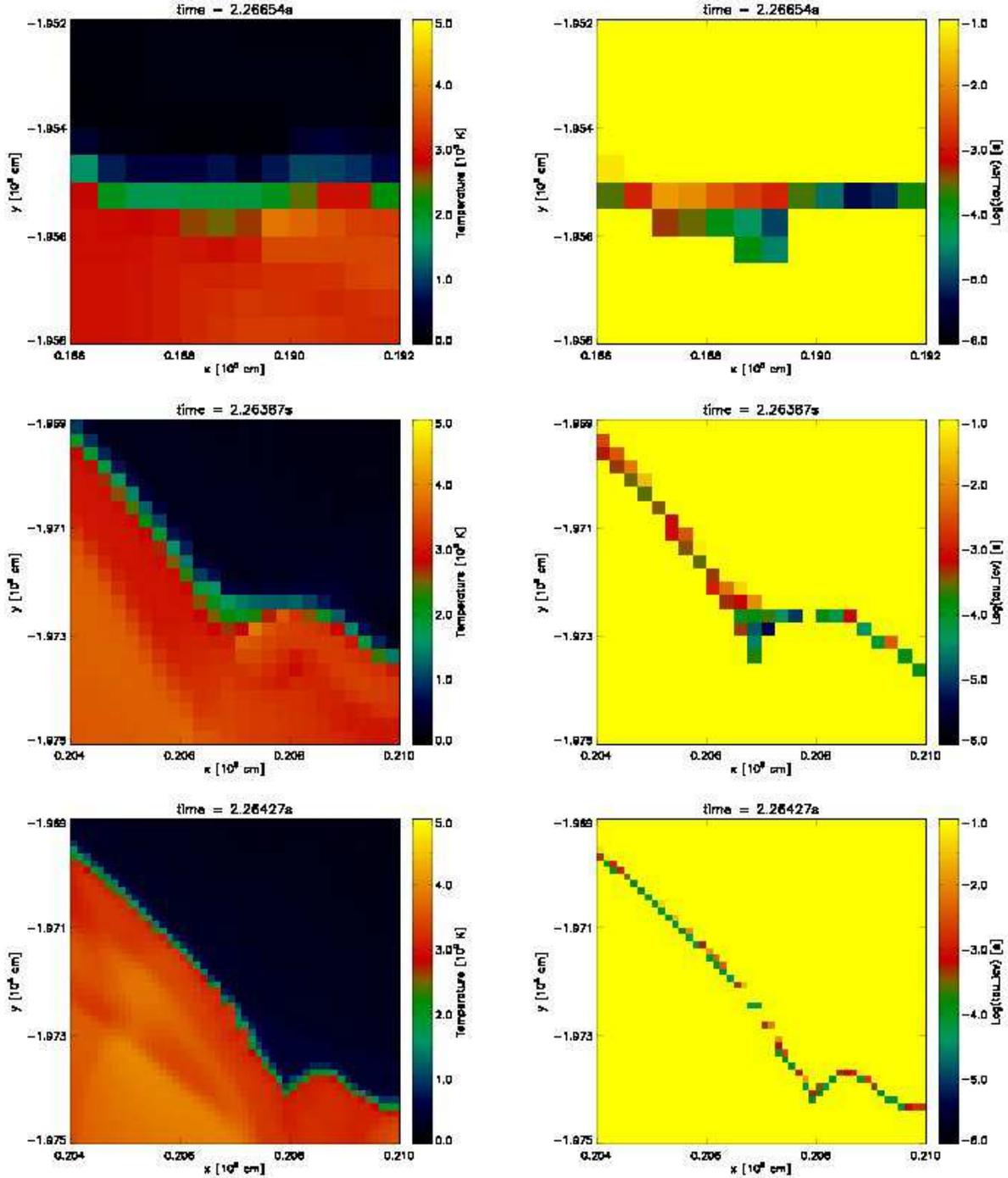}
\caption{Temperature (left) and corresponding logarithm of the constant volume induction time $\tau_{i,cv}$ (right) for 500 m, 250 m, and 125 m resolution (top to bottom). 
The sidelength of each plot is 6 km. 
 $\tau_{i,cv}$ (see eq.~\ref{eq:tauicv}) was set to 1 s for $X_0^c > 0.9$. For 500 m and 250 m the time of the snapshots is just before the detonation shock forms. 
For 125 m the time of the snapshots is just after the compression wave has passed the feature near $x = 208$ km.
\label{fig:tiny}} 
\end{figure}

\begin{figure}
\plotone{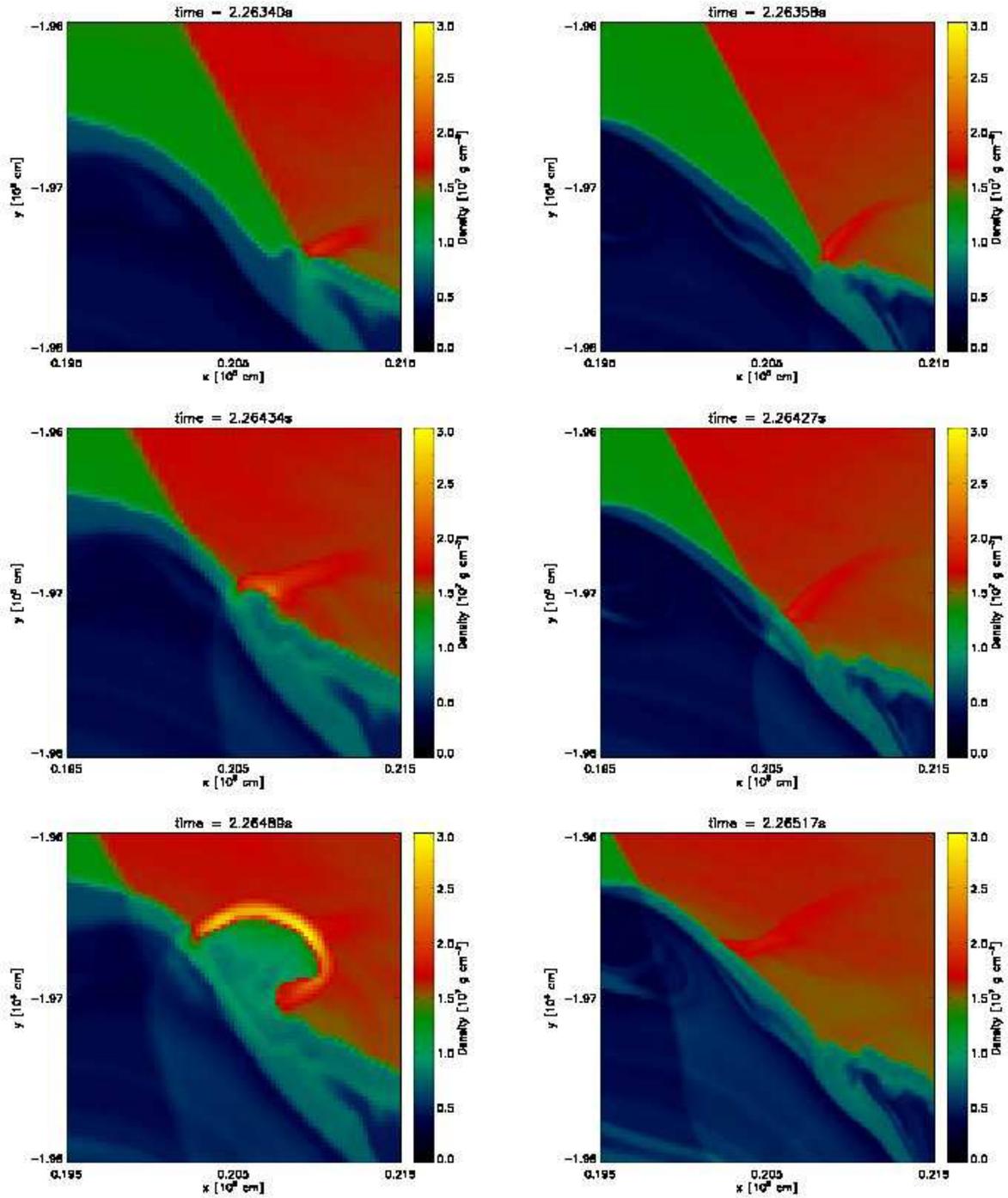}
\caption{Shown in each column is a time sequence of the density.  Left: 250 m resolution. Right: 125 m resolution. \label{fig:dens_small}} 
\end{figure}

\begin{figure}
\plotone{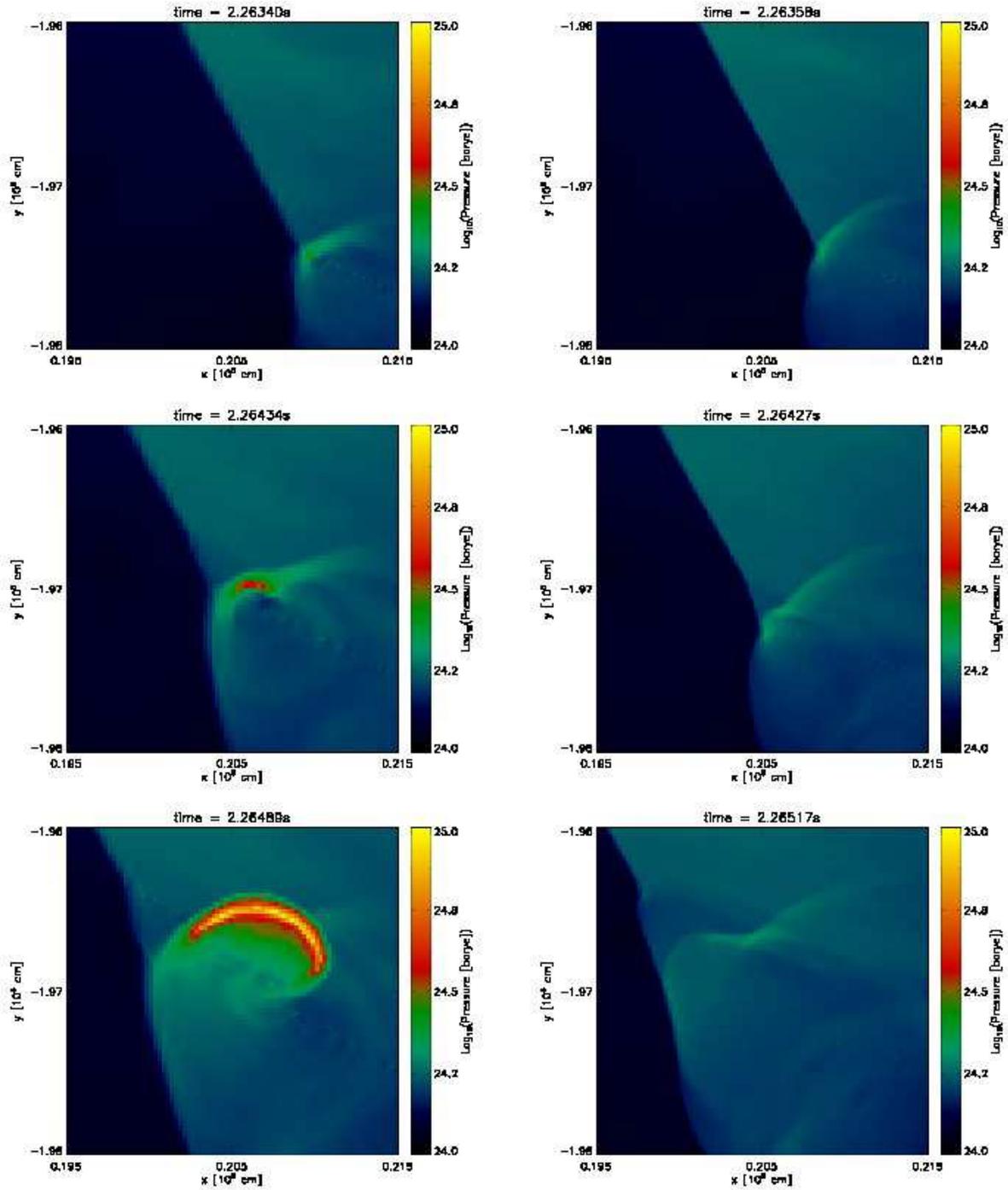}
\caption{Shown in each column is a time sequence of the pressure.  Left: 250 m resolution. Right: 125 m resolution. \label{fig:pres_small}} 
\end{figure}

\begin{figure}
\plotone{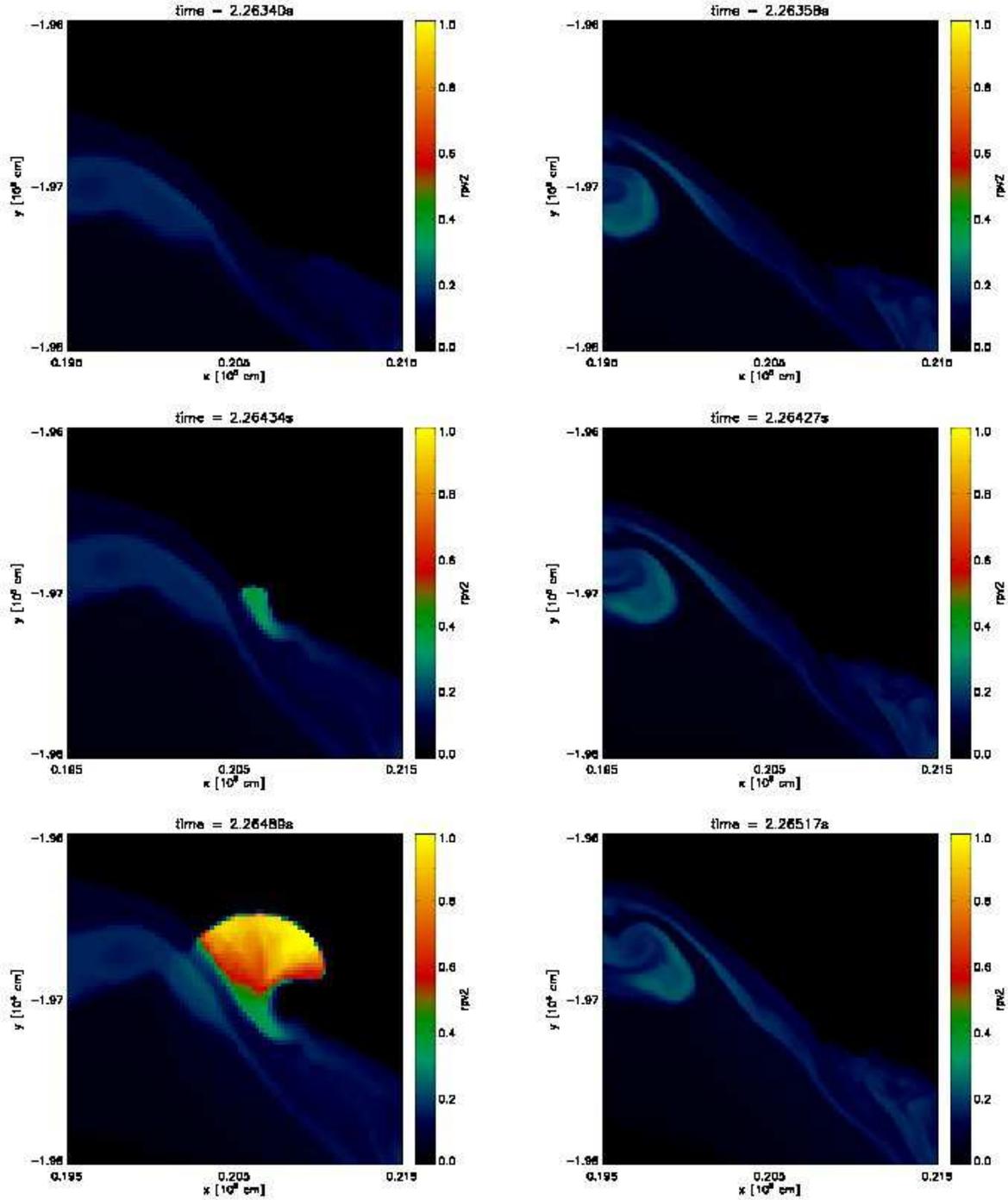}
\caption{ Shown in each column is a time sequence of the reaction progress variable $\phi_2$, a tracer for the progress of the burning to Si-group elements \citep{townsley07}. Left: 250 m resolution. Right: 125 m resolution. \label{fig:rpv2_small}} 
\end{figure}

In this case, the detonation forms off axis when the leading edge of the jet has advanced to a density of $\sim\rho=1.2\times10^7$ g cm$^{-3}$ (see Table \ref{tab:conditions} and Fig.~\ref{fig:dens_small}) in a location close to where the 500 m case detonated.  
The detonation occurs only $\approx 0.002$ s earlier than in the 500 m case.  
While both the 500 m and the 250 m simulations rely on the moving compression front for initiation, and detonate in a similar location, the mechanism is slightly different in the two cases. 
The 250 m model detonates when the temperature gradient at the head of the jet is broadened over a few zones extending into unburned fuel (see Fig. \ref{fig:tiny} middle left). 
The shallow temperature gradient is established when a compression front and an associated reflected wave interact with a Kelvin-Helmholtz cusp (see \ref{fig:dens_small}), which is not present in the lower resolution simulations. 
The high temperature extends into a region of unprocessed fuel and a relatively shallow gradient in the isochoric (constant volume) induction time is set up (see Fig. \ref{fig:tiny} middle right). 
The coherent energy release due to carbon burning leads to the build up of a shock with high enough pressure (see fig.~\ref{fig:pres_small}) that in its wake oxygen burning commences along a line along the original induction time gradient (see Fig.~\ref{fig:rpv2_small}). Once oxygen has ignited, the detonation begins to propagate radially outward from the initiation site.

\subsection{125 m resolution case}
This case initially resembles the 250 m case. 
The same hydrodynamical wave that triggered detonation in the 250 m case crosses the same Kelvin-Helmholtz pocket of unburned fuel (see Fig. \ref{fig:dens_small}). 
Unlike in the 250 m case, however, oxygen fails to ignite (that is, $\phi_2 \longrightarrow \!\!\!\!\!\!\!\!/\;\;\; 1$, see Fig. \ref{fig:rpv2_small}), and the high pressure spot that leads to a detonation in the lower resolution case eventually dissipates without initiating a detonation (see Fig. \ref{fig:pres_small}). 
Thus in this case (which is the highest resolution simulation we were able to carry out), the compression wave leads to increased carbon burning in the region, but does not trigger an outburst of oxygen burning and is not sufficient to cause initiation of a detonation.  This may be due to the fact that the temperature gradients, and therefore the induction time gradients, are much steeper, even in the cusp, compared to the lower resolution case (see fig. \ref{fig:tiny} bottom). 

Unfortunately, the computational demand of this simulation was great enough that we were unable to continue it to later times.  At the time we stopped the simulation, the burning front at the head of the jet is still moving toward higher densities, and since the critical length scale for the initiation of a detonation is a strongly decreasing function of the density (see table 4 from \citet{seitenzahl09b}), a detonation may occur ahead of the jet at a later time.  In addition, the strong sound wave that is involved in triggering the detonation in the 500 m and 250 m cases is still moving toward the symmetry axis, which will increase the compression in the wave as it approaches the axis, making a detonation more likely.


\section{Discussion}
\label{sec:dis}

\subsection{Comparison with Earlier Work}

\citet{plewa07} states that in his GCD models the detonation always forms via a shock to detonation transition (SDT) \citep{bdzil92,sharpe02}. 
In our results, the initiation of the detonation is not tied to a shock. 
The 500 m and 250 m cases are somewhat similar in spirit to an SDT, since they also rely on the preconditioning of fuel by a hydrodynamic wave. 
However, the wave is not a shock and the flows leading up to detonation are not supersonic.
While both simulations were performed using \FLASH\, there are significant differences in the employed flame models, energetics, and initial conditions, which could account for the disagreement. 
Another difference is the 8 km resolution employed in the \citet{plewa07} study, which is 2-64 times coarser than in our study.

\subsection{Initiation of a Detonation in the GCD Model}

Initiation of a detonation occurs in three of the fours simulations we carried out.  In all three cases, detonation occurs via the gradient mechanism. We consider the detonation in the 4 km case to be an artifact of the low resolution and numerical diffusion, which lead to a much shallower temperature gradient, and hence larger pre-heated region, than in the higher resolution cases.  In the other two cases in which a  detonation occurs, it is due to the passage of a compression front (500 m) or the passage of a compression front followed by a reflected wave (250 m) through the temperature gradient ahead of the jet, broadening the temperature gradient and making it shallow enough (and therefore the induction time long enough) that the gradient mechanism can operate.  These mark the first simulations of gradient-induced detonations in a whole star simulation.

In the fourth case (125 m), the highest resolution simulation we carried out, initiation of a detonation had not yet occurred by the time we stopped the simulation because of the computational demand it required.
Whether a detonation eventually occurs in this case is an open question. We are therefore unable to state that initiation of a detonation via the gradient mechanism occurs in all of the higher resolution simulations.  

Our studies confirm that the temperatures and densities needed for initiation of a detonation to occur are readily obtained.  The primary challenge is the necessity of pre-conditioning the fuel so that a shallow enough temperature gradient (and therefore a large enough induction time) occurs.  Sufficiently shallow temperature gradients exist in the jet, but the material within the jet is mostly ash, due to the highly reactive nature of the fuel.  Consequently, it is necessary that a sufficiently shallow temperature gradient occur in the unburned fuel ahead of the jet.  For the resolutions explored in this paper, the spatial extent of the temperature gradient decreases as the resolution becomes finer, and the temperature gradient steepens with increasing resolution.  It is therefore not too surprising that the 125 m case (the highest resolution simulation we carried out) fails to detonate at the location where a detonation occurs in the 500 m and 250 m cases, or anywhere else (at least until the time the evolution was stopped).  This is somewhat disconcerting.

\begin{figure}
\plotone{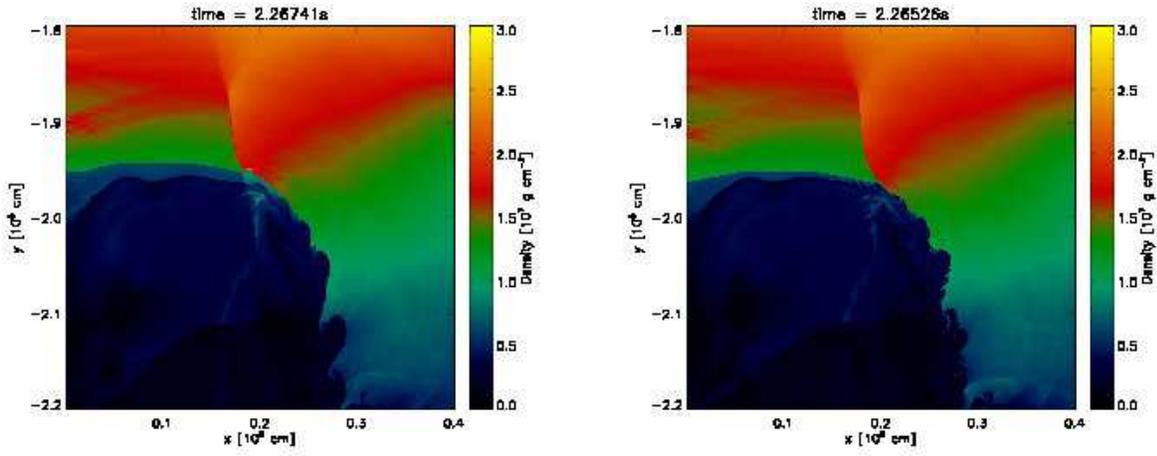}
\caption{Comparison of density in the jet for 500 m resolution (left) and 125 m resolution (right). While the 125 m case shows turbulent structure on smaller scales, there is overall qualitative agreement. \label{fig:res_comp}} 
\end{figure}

However, apart from the fact that an admixture of helium \citep[e.g.][]{yoon05} would lead more robustly to a detonation (see Table 12 in \citet{seitenzahl09b}), there are several reasons that suggest a detonation might still occur in the highest resolution case. 
First, the hot jet is still moving toward higher densities, and since the critical length scale for the initiation of a detonation is a strongly decreasing function of density (see Table 4 in \citet{seitenzahl09b}), a detonation may still occur ahead of the jet at a later time. 
Second, the strong sound wave that is involved in triggering the detonation in the 500 m and 250 m cases is still moving toward the symmetry axis, which will increase the compression in the wave as it approaches the axis, making a detonation more likely.
Third, the higher the resolution, the more turbulent shear flow structures emerge on small scales at the jet-fuel boundary (see fig. \ref{fig:res_comp}). 
Before the high resolution truncated cone was imposed at $t=2.2124$ s, the resolution of the simulation was 4 km, a scale at which Kelvin-Helmholtz shear instabilities are effectively suppressed.
After the resolution in the truncated cone is suddenly increased, there is no structure initially on scales $< 4$ km.  The turbulence on  smaller scales requires time to develop; if given time to fully mature, it will smear the fuel-ash interface, leading to shallower induction time gradients, which will facilitate initiation of a detonation. 

In this regard, we note that \citet{knystautas79} and \citet{thomas00} list three criteria that need to be fulfilled for a transition to a detonation in the turbulent shear layer of a jet:  (1) large-scale energetic eddies in the unburned gas (2) strong fine turbulence sufficient to efficiently mix hot ash into the fuel; and (3) the establishment of an induction time gradient in the mixed region such that the SWACER mechanism can act.  All three of these criteria appear to be fulfilled in our case.

\section{Conclusion}
\label{sec:con}

The high-resolution simulations reported in this paper suggest that the detonations in lower resolution simulations (i.e., the 4 km case in this paper and the lower resolution simulations in earlier studies) are largely numerical.  Two high-resolution simulations (500 m and 250 m) demonstrate that pre-conditioning of a fuel pocket surrounded by Kelvin-Helmholtz billows by a compression wave moving through the pocket can lead to a gradient initiated detonation.  For the resolutions explored in this paper, the spatial extent of the temperature gradient decreases as the resolution becomes finer, causing the temperature gradient to steepen with increasing resolution.  Consequently, the temperature gradient becomes too steep (and therefore the induction time becomes too short) in the highest resolution simulation (125 m) for a detonation to occur at the time and location at which it occurs in the 500 m and 250 m simulations, and did not occur by the time we had to stop the simulation because of its computational demand.  However, it may detonate later.  In addition, shear-induced turbulence, which surrounds the jet and is largely suppressed at low resolution, may broaden the temperature gradient, leading to conditions favorable for the initiation of a detonation, especially if the influence of compression waves on the turbulent layer is considered.  
This suggests that future studies should be done (ideally in 3D) at even higher resolution, beginning earlier and running for longer times, in order to capture the development of the turbulence surrounding the jet more self consistently.

\acknowledgments
We would like to thank Dean Townsley and George C. Jordan IV for their helpful suggestions and discussions. This work is supported in part by the U.S. Department of Energy under contract B523820 to the ASC Alliances Center for Astrophysical Flashes and in part by the National Science Foundation under grant PHY 02-16783 for the Frontier Center "Joint Institute for Nuclear Astrophysics" (JINA) and in part by the Emmy Noether Program of the German Research Foundation (DFG; RO~3676/1-1). JWT acknowledges support from Argonne National Laboratory, operated under contract No. W-31-109-ENG-38 with the DOE. 


\clearpage

\clearpage

\clearpage

\clearpage

\clearpage

\clearpage

\clearpage

\clearpage

\clearpage

\clearpage

\clearpage

\clearpage

\clearpage
\begin{deluxetable}{rccccc}
\tabletypesize{\footnotesize}
\tablewidth{0pc}
\tablecaption{Detonation conditions \label{tab:conditions}}
\tablehead{ \colhead{Resolution} & \colhead{Time}   & \colhead{Location (x,y)} & \colhead{Fuel Density} & \colhead{$T_{max}$} & \colhead{$T_0$} \\  \colhead{[$10^2$ cm]} &\colhead{[s]}  &    \colhead{  [$10^8$ cm]}  & \colhead{[$10^7$ g cm$^{-3}$]} & \colhead{[$10^9$ K]} & \colhead{[$10^7$ K]} } 
\startdata
4,000    & 2.204 & (0.00,-2.04) & 1.1 & $\sim$3.0 & $\sim$10 \\
   500    & 2.266 & (0.19,-1.95) & 1.2|1.6 & $\sim$3.0 & $\sim$20  \\
   250   &  2.264 & (0.21,-1.97)  & 1.2|1.6 & $\sim$3.0 & $\sim$20 \\
   125 &  N/A & N/A & N/A &N/A & N/A
\enddata
\end{deluxetable}

\end{document}